\documentclass{article}
\usepackage{FPSpro}
\newcommand{\mic}   { \mu{\mathrm m} }

\newcommand{\Vub}   { {\mathrm V}_{\mathrm{ub}} }
\newcommand{\Vcb}   { {\mathrm V}_{\mathrm{cb}} }
\newcommand{\Vtd}   { {\mathrm V}_{\mathrm{td}} }
\newcommand{\Vts}   { {\mathrm V}_{\mathrm{ts}} }
\newcommand{\aVub}  { |{\mathrm V}_{\mathrm{ub}}| }
\newcommand{\aVcb}  { |{\mathrm V}_{\mathrm{cb}}| }
\newcommand{\aVtd}  { |{\mathrm V}_{\mathrm{td}}| }
\newcommand{\aVts}  { |{\mathrm V}_{\mathrm{ts}}| }
\newcommand{\ups}   { \Upsilon{\mathrm{(4S)}} }
\newcommand{\Bplus} { {\mathrm B}^+ }
\newcommand{\Bzero} { {\mathrm B}^0 }
\newcommand{\Bzbar} { \overline{{\mathrm B}^0 }}
\newcommand{\Bs}    { {\mathrm B}^0_{\mathrm s} }
\newcommand{\Bq}    { {\mathrm B}^0_{\mathrm q} }
\newcommand{\Bqbar} { \overline{{\mathrm B}^0_{\mathrm q }}}
\newcommand{\Ds}    { {\mathrm D}_{\mathrm s} }
\newcommand{\lamb}  { \Lambda_{\mathrm b} }
\newcommand{\B}     { {\mathrm B} }
\newcommand{\Dstar} { {\mathrm D}^\star }
\newcommand{\bl}    { {\mathrm b}\to {\mathrm X} \ell \nu }
\newcommand{\blc}   { {\mathrm b}\to {\mathrm X}_{\mathrm c} \ell \nu }
\newcommand{\blu}   { {\mathrm b}\to {\mathrm X}_{\mathrm u} \ell \nu }
\newcommand{\bcl}   { {\mathrm b}\to {\mathrm c}\to  {\mathrm X} \ell \nu }
\newcommand{\BRbl}  { {\mathrm{BR}}( {\mathrm b}\to {\mathrm X} \ell \nu )}
\newcommand{\BRblc} { {\mathrm{BR}}( {\mathrm b}\to {\mathrm X}_{\mathrm c} \ell \nu) }
\newcommand{\BRblu} { {\mathrm{BR}}( {\mathrm b}\to {\mathrm X}_{\mathrm u} \ell \nu) }

\newcommand{\BRBl}  { {\mathrm{BR}}( {\mathrm B}\to {\mathrm X} \ell \nu )}
\newcommand{\Dmd}   { \Delta m_{\mathrm d} }
\newcommand{\Dms}   { \Delta m_{\mathrm s} }
\newcommand{\Dmq}   { \Delta m_{\mathrm q} }
\newcommand{\ips}   { {\mathrm {ps}}^{-1} }
\newcommand{\gevc}  { {\mathrm {GeV}}\!/\!c }
\newcommand{\amp}   { {\cal{A}} }

\newcommand{\eg}    {{\it{e.g.}}}
\newcommand{\etal}  {{\it{et al.}}}
\begin{document}
\title{ REVIEW OF B PHYSICS RESULTS\\ FROM THE LEP EXPERIMENTS AND SLD
  }
\author{
  Duccio Abbaneo        \\
  {\em CERN, EP Division, CH-1211, Geneva 23} 
  }
\maketitle
\baselineskip=11.6pt
\begin{abstract}
A review of b physics results from the LEP experiments and SLD is presented.
Emphasis is given to the determinations of the $\aVcb$ and $\aVub$,
and to the study of B meson oscillations, which yield
bounds on the unitarity triangle.
\end{abstract}
\baselineskip=14pt
\section{Introduction}
Over the past decade, many important b physics measurements were performed
at Z factories (LEP, SLC). The large boost of b hadrons gives access to 
important properties (\eg~lifetimes, 
oscillation frequencies of neutral B mesons),
which cannot be studied with b hadrons at rest.
Today asymmetric B factories are taking over for what concerns the physics of
$\Bzero$ and $\Bplus$ mesons, while our experimental 
knowledge of $\Bs$ and b baryon physics is still based on
measurements performed by the LEP experiments and  SLD,  and by CDF at the
Tevatron. A major step forward in these topics will be made only when 
significant statistics from the Tevatron Run II are analyzed.
In this review emphasis is given to studies of b decay properties 
rather than b hadron production; in particular to measurements that
are related to the determination of CKM matrix elements, and hence
to the description of CP violation in the 
Standard Model\footnote{For some topics, new measurements
have been released in the weeks following the conference. The
results presented in this report include all analyses available
at the end of July 2001.}.

First, results on b hadron lifetimes are reviewed. The inclusive
b lifetime is an input parameter for the derivation of $\aVcb$
and $\aVub$ from the inclusive semileptonic branching ratios,
while individual b-hadron lifetimes provide an important test
of our understanding of hadron dynamics. The lifetime difference
in the $\Bs$ system is now also measured; the ratio to the oscillation
frequency can be calculated on the lattice.
Next, measurements of inclusive semileptonic b decay rates
are presented, which give the opportunity to derive 
 $\aVcb$ and $\aVub$. An alternative determination of $\aVcb$
is provided by the study of $\B \to \Dstar \ell \nu$ exclusive decays.
Finally, studies of neutral B meson oscillations yield information on
$\aVtd$ and $\aVts$.

Most of the results presented are based on analyses from the LEP experiments
and SLD. The LEP I data sample consists of almost four million hadronic
decays per experiment. SLD has collected a sample 10 times
smaller than that of each LEP experiment, but is competitive on some
specific analyses, due to some unique features
of the accelerator and the detector. 
The polarization of the electron beam, the tiny and stable beam spot and the
excellent precision of the CCD vertex detector significantly enhance the 
quark charge tagging capability, the precision in the track impact paremeter
and b decay length measurements, and the efficiency in inclusive secondary vertex finding.

\section{Measurements of b hadron lifetimes}
\label{sec.life}
Measurements of the inclusive b lifetime are based on semileptonic
or fully inclusive final states, either measuring the impact parameter
of charged particle tracks, or the decay length of inclusively reconstructed secondary
vertices. 

\begin{figure}[t]
  \vspace{4.8cm}
  \includegraphics{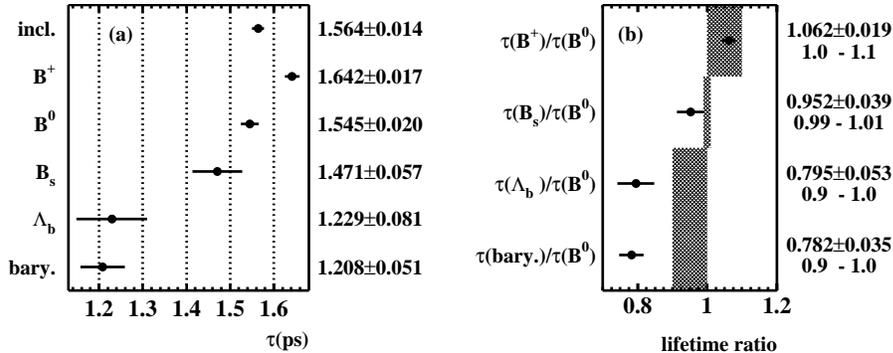}
  \caption{\it
    (a) Experimental results for inclusive and individual b hadron lifetimes. The averages include all analyses from CDF, SLD and the LEP experiments.
    (b) Comparison of lifetime ratios with theoretical predictions.   
    \label{fig.life} }
\end{figure}

Individual b hadron lifetimes are measured with several techniques.
The most precise results come from reconstruction of semileptonic final
states and from inclusive reconstruction of secondary vertices, particularly
suitable for $\Bplus$ mesons. 
The averages of results from CDF, SLD and the LEP experiments
are shown in Fig.~\ref{fig.life}a; the ratio
of individual lifetimes to the $\Bzero$ lifetimes are compared with
theoretical predictions\cite{bigi} in Fig.~\ref{fig.life}b.
The predicted hyerarchy is observed. The difference between the 
$\Bzero$ and $\Bplus$ lifetimes is established experimentally at more
than $3\sigma$. A discrepancy between measurement and prediction of about $3\sigma$
emerges for the b baryon lifetime.

New measurements of $\Bzero$ and $\Bplus$ lifetimes are 
now coming from the asymmetric B factories:  the precision of the results presented here 
is expected to be exceeded by the end of 2001. For $\Bs$ and b baryon lifetimes,
no significant improvement is to be expected from new analyses of the available
data samples. 

The discrepancy between theory and experiments  for the b baryon lifetime has triggered new theoretical studies. 
Neubert\cite{neub_life} has performed an analysis with less modelling assumptions, 
varying the unknown hadronic matrix elements within plausible ranges. The predictivity
for the hyerarchy between $\Bzero$ and $\Bplus$ is lost, and the low measured
value of the baryon lifetime is still difficult to accommodate. 
Analyses based on  QCD sum rules\cite{fulvia} also give a prediction higher than
the measurement, although Huang~\etal\cite{huang_life}, by stretching some
assumptions, have been able to produce a low value, 
$\tau(\lamb) / \tau(\Bzero) = 0.86 \pm 0.04$,
in better agreement with the experimental determination.

\begin{figure}[h]
 \vspace{5.0cm}
  \includegraphics{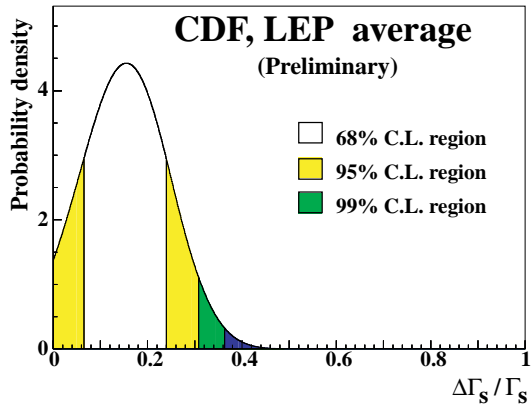}
  \caption{\it
    {Probability density function for $\Delta \Gamma_{\mathrm s} / \Gamma_{\mathrm s}$ from all
available analyses, with the additional constraint $\Gamma_{\mathrm s} = \Gamma_{\mathrm d}$.}
    \label{fig.dgs} }
\end{figure}

Several methods have been explored by the LEP experiments and CDF to constrain the
lifetime difference in the $\Bs$ system. The simplest analyses are based on the 
observation that fitting for the 
$\Bs$ lifetime in inclusive samples, 
or samples of semileptonic $\Bs$ decays, 
yields a result which takes a second
order correction from a possible lifetime difference bewteen 
the two $\Bs$ states.
Assuming that the $\Bs$ and $\Bzero$ decay widths are equal
($\Gamma_{\mathrm s} = \Gamma_{\mathrm d}$), 
the result
can be translated to a constraint on the lifetime difference. Alternatively, 
the selection can be aimed at enhancing the $\Bs$ short content of the sample analysed. 
In this case the fraction of $\Bs$ short has to be evaluated, and the sensitivity 
of the fitted lifetime to the lifetime difference is higher. Finally, ALEPH tried to 
select $\Ds^+ \Ds^-$ final
states, corresponding to $\Bs$ short decays, by selecting jets containing two reconstructed
$\phi$ mesons. Fitting for the lifetime of this sample yields a direct 
measurement of the $\Bs$ short lifetime. In addition, since this final state
is the only significant contribution to the width difference, a measurement
of the rate can also be translated to a constraint on the lifetime difference.
All these methods yield rather mild constraints on the $\Bs$ lifetime difference,
but combining the likelihood profiles from all the analyses,
together with the contraint $\Gamma_{\mathrm s} = \Gamma_{\mathrm d}$,  gives the result reported
in Fig.~\ref{fig.dgs}, which can be quantified either as a measurement or as a 95\% C.L. limit:
\begin{equation}
\Delta \Gamma_{\mathrm s} / \Gamma_{\mathrm s}  = 0.16\, ^{+0.08}_{-0.09} \ , \ \ \ \ \ \ \ \ \ 
\Delta \Gamma_{\mathrm s} / \Gamma_{\mathrm s}  < 0.31\, @\ 95\%\ {\mathrm{C.L.}} \ ,
\end{equation}
in good agreement with the theoretical prediction\cite{beneke_dgs} of $\Delta \Gamma_{\mathrm s} / \Gamma_{\mathrm s}  = 0.097\, ^{+0.038}_{-0.050}$. 

\section{Semileptonic decays}
The inclusive direct semileptonic decay rate of b hadrons $\BRbl$ 
is measured by the LEP experiments.
High-purity b hadron samples are obtained
by applying a lifetime tagging in the opposite event-half, while
a reliable knowledge of the lepton identification and background is achieved
thanks to several control samples which allow the simulation to be precisely
tuned with the data. The challenge of these analyses is to disentangle the various
sources of prompt leptons in b decays, in particular direct $\bl$ from cascade $\bcl$ decays,
keeping control of the related systematic uncertainties.

\begin{figure}[t]
  \vspace{5.6cm}  \includegraphics{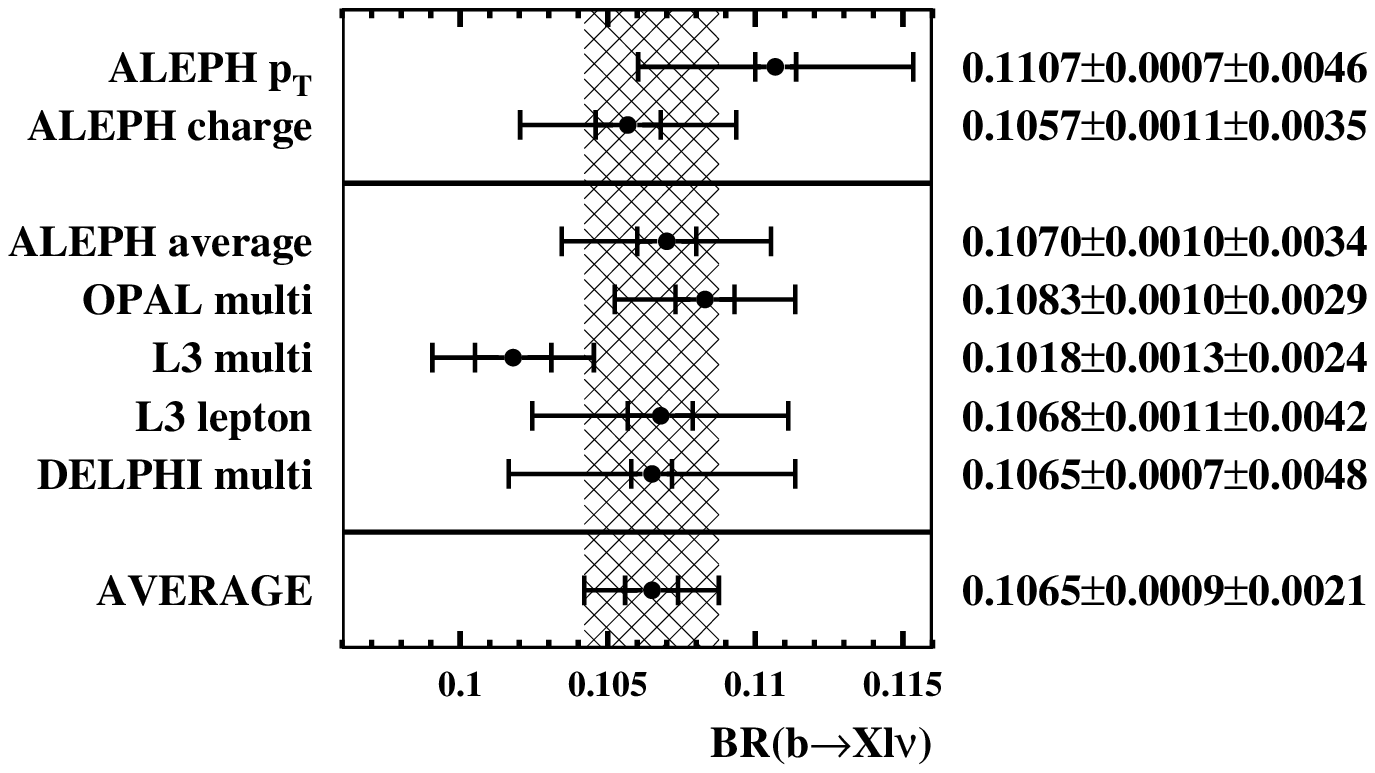}
  \caption{\it
    Measurements of $\BRbl$ at LEP, and combined value.
    \label{fig.btol} }
\end{figure}

The measurements available are reported in Fig.~\ref{fig.btol}, together with the average.
Two new results from ALEPH have been recently released, obtained with methods that have
largely independent systematic uncertainties.

The $\BRbl$ value, together with the measurement of the inclusive b lifetime,
yield a determination of the $\aVcb$\cite{bigi}, once
the small $\blu$ contribution has been subtracted,
\begin{equation}
\aVcb = (41.1\pm 2.5)\times 10^{-3} \times \sqrt{ \frac{ \BRblc }{0.105}}
\times \sqrt{ \frac{1.55\ {\mathrm{ps}}}{\tau_{\mathrm{b}}^{\mathrm{incl}}}} \ ,
\label{eq.vcb} 
\end{equation}
obtaining 
\begin{equation}
\aVcb = (40.9 \ \pm\, 0.5_{\mathrm{exp}}\ \pm\, 2.4_{\mathrm{theo}})\times 10^{-3}\ , 
\label{resu.vcb} 
\end{equation}
where the uncertainty is dominated by the theoretical error from Eq.~\ref{eq.vcb}.
The $\BRbl$ result from LEP can also be compared with the $\ups$ value, after
correcting for the different b hadron mixture. Assuming equal semileptonic widths
for all b species, the correction can be written as
\begin{equation}
{\left. \BRBl \right|}_{\mathrm{LEP}}= \BRbl \times \frac{\tau_{\mathrm B}}{\tau_{\mathrm b}} = 0.1085 \pm 0.0027 \ ,
\end{equation} 
which compares rather well with the value measured by CLEO\cite{PDG2000} 
\begin{equation}
{\left. \BRBl \right|}_{\mathrm{CLEO}}= 0.1049 \pm 0.0046\ .
\end{equation} 

\begin{figure}[t]
  \vspace{4.1cm}  \includegraphics{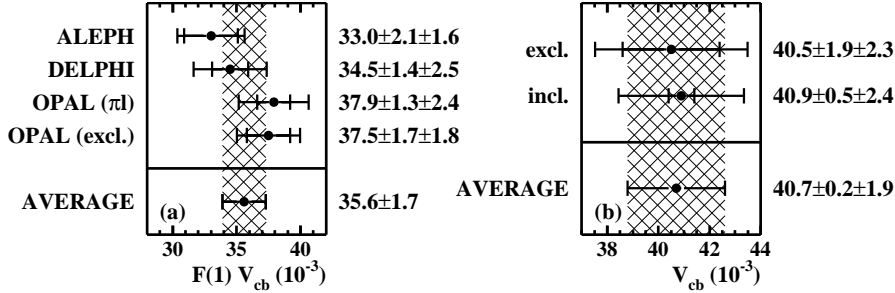}
  \caption{\it
    (a) Measurements of  ${\cal{F}}(1) |\Vcb|$,$\BRbl$ at LEP, and combined value.
    (b) Inclusive and exclusive determination of  $|\Vcb|$, along with their combination; the
    first error quoted comes from the uncertainties on the experimental inputs, the second
    from theory.
    \label{fig.vcb} }
\end{figure}

An alternative determination of $\aVcb$ is provided by the study of
exclusive $\B\to \Dstar \ell \nu$ decays. The decay rate can be written
as a function of the $\Dstar$ boost in the $\B$ rest frame $\omega$, as\cite{vcb_excl}
\begin{equation}
\frac{d\Gamma}{d\omega} \propto {\cal{K}}(\omega) {\cal{F}}^2(\omega) \aVcb^2 \ ,
\end{equation}
where ${\cal{K}}(\omega)$ is a phase space factor and ${\cal{F}} (\omega)$ is
an unknown hadronic form factor. In the infinite b quark mass limit, the hadronic
form factor equals unity for a $\Dstar$ at rest ($\omega = 1$).
Mass and non-perturbative corrections can be calculated with the Operator
Product Expansion formalism\cite{bigi},
obtaining ${\cal{F}}(1) = 0.88 \pm 0.05$. In the analyses the 
spectrum of the candidates as a function of the reconstucted $\omega$ is
fit for the slope $\rho^2$ and the intecept at $\omega = 1$, ${\cal{F}}(1) |\Vcb|$,
where the phase space vanishes. The results are averaged accounting for the
correlation between the two free parameters, obtaining the results shown in
Fig.~\ref{fig.vcb}a.
The determintation from the exclusive channel is in perfect agreement with that
from the inclusive $\bl$ rate, as shown in Fig.\ref{fig.vcb}b, where a global
average is also presented.

\begin{figure}[t]
  \vspace{3.7cm}  \includegraphics{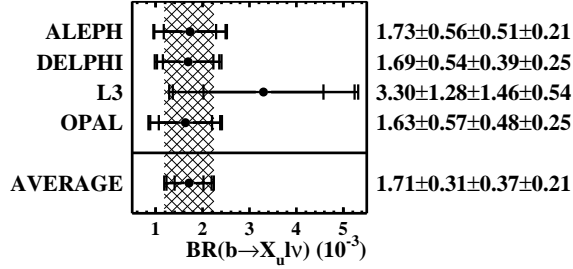}
  \caption{\it
    Measurements of $\BRblu$ at LEP, and combined value. The first uncertainty
    quoted accounts for limited statistics and detector effects, the second
    for the modelling of the $\blc$ background and the third for the 
    modelling of the $\blu$ signal.
    \label{fig.bul} }
\end{figure}

The LEP experiments have also performed analyses of inclusive
semileptonic decays, where charmed and charmless hadronic final states
are discriminated on a statistical basis, producing measurements
$\BRblu$. Discriminating variables are aimed at
selecting low-mass hadronic states; the task is complicated by the
high multiplicity of fragmentation particles, which need to be
disentangled from the b decay products. The analyses try to achieve
the best possible separation between charmed and charmless final states
by combining several discriminating variables by means of neural
networks. Nevertheless the signal is measured on top of a large
background from $\blc$ transitions, which need to be subtracted;
the estimate of the related systematic uncertainties is the critical 
issue for these analyses. On the other hand, because the b hadron
has a large boost, the selection has relatively small dependence
upon the decay kinematics, and therefore upon the modelling of the signal.
The results available and the LEP average are shown in Fig.~\ref{fig.bul}.

Similarly to Eq.~\ref{eq.vcb}, the measured branching ratio can be used
to extract the $\aVub$\cite{ural} according to
\begin{equation}
\aVub = (4.45\pm 0.18)\times 10^{-3} \times \sqrt{ \frac{ \BRblu }{0.002}}
\times \sqrt{ \frac{1.55\ {\mathrm{ps}}}{\tau_{\mathrm{b}}^{\mathrm{incl}}}} \ ,
\label{eq.vub} 
\end{equation}
which yields
\begin{equation}
\aVub = (4.09 \ 
{^{+0.36}_{-0.39}}_{\mathrm{stat+exp}}\ 
{^{+0.42}_{-0.47}}_{\mathrm{b\to c}}\ 
{^{+0.24}_{-0.26}}_{\mathrm{b\to u}}\ 
\pm\, 0.17_{\mathrm{theo}})\times 10^{-3}\ , 
\label{resu.vub} 
\end{equation}
where the first error accounts for limited statistics and
detector effects, the second and third come from the modelling of $\blc$ and $\blu$ transitions,
respectively, and the last reflects the uncertainty in Eq.~\ref{eq.vub}. 

\section{Neutral B meson oscillations}
The oscillation frequency in the $\Bzero - \Bzbar$ system,
which is proportional to the mass difference of the two
eigenstates, can be translated to a measurement of $ \aVtd $,
$ \Dmd \propto {  \aVtd  }^2 \times ({\mathrm{QCD\ corrections}})$, 
yielding a constraint on the size of the CP violating phase $\eta$.
Unfortunately QCD effects are large and the associated 
uncertainty dominates the extraction of  $\Vtd$.
A better constraint on $\eta$ could be obtained from the ratio of
the oscillation frequencies of $\Bs$ and $\Bzero$ mesons, since some
of the QCD uncertainties cancel in the ratio. The factor
$\xi$ in Eq.~\ref{eq:ratio} is estimated to be known at the $5\%$ level.
\begin{equation}
  \frac{\Dms}{\Dmd} = 
  \frac{m_{\Bs}}{m_{\Bzero}} \xi^2 {\left| \frac{\Vts}{\Vtd} \right|}^2 \ .
  \label{eq:ratio}
\end{equation}
The
proper time distributions of ``mixed'' and ``unmixed'' decays, given in Eq.~\ref{eq:pdf},
are measured experimentally.
The oscillating term introduces a time-dependent difference between the two classes.
\begin{eqnarray}
  {\cal P}(t)_{\Bq \to \Bqbar} & = & 
  \frac{\Gamma e^{-\Gamma t}}{2} \, [1 - \cos (\Dmq \, t)] \ ,
  \nonumber \\
  {\cal P}(t)_{\Bq \to \Bq} & = & 
  \frac{\Gamma e^{-\Gamma t}}{2} \, [1 + \cos (\Dmq \, t)] \ , 
  \label{eq:pdf}
\end{eqnarray}
The amplitude of such difference is damped not only by the 
natural exponential decay,
but also by the effect of the experimental resolution in the proper time
determination. The proper time is derived from the measured decay length
and the reconstructed momentum of the decaying meson. The resolution on 
the decay length
$\sigma_L$
is to first order independent of the decay length itself, and is largely determined
by the tracking capabilities of the detector. The momentum resolution $\sigma_p$
depends strongly on 
the final state chosen for a given analysis, and is typically proportional to the momentum 
itself. 
\begin{figure}[t]
  \vspace{6.9cm}
  \includegraphics{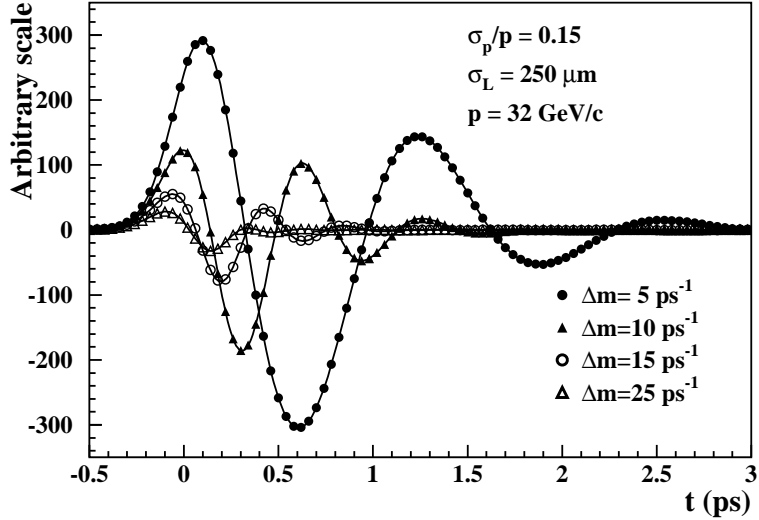}
  \caption{\it
    Difference in the proper time distributions of unmixed and mixed decays
    for monochromatic $\B$ mesons, fixed
    decay length and momentum resolutions, and different values of the
    oscillation frequency.
    \label{fig:time} }
\end{figure}
The proper time resolution can be therefore written as:
\begin{equation}
  \sigma_t = \frac{m}{p} \sigma_L \oplus \frac{\sigma_p}{p} \, t \ ,
  \label{eq:timeres}
\end{equation}
where the decay length resolution contributes a constant term, and the momentum resolution
a term proportional to the proper time. Examples of the resulting 
observable difference are 
shown in Fig.~\ref{fig:time}, for the simple case of monochromatic $\B$ mesons of 
momentum $32 \ \gevc$, resolutions of $\sigma_p/p = 0.15$ and $\sigma_L =250\
\mic$ (Gaussian), and for different values of the true oscillation frequency.
For low frequency several periods can be observed. As the frequency 
increases, the
effect of the finite proper time resolution becomes more relevant, 
inducing an overall decrease of observed difference, and a faster damping as a function
of time (due to the momentum resolution component). 
In the example given, for a frequency
of $25\ \ips$ only a small effect corresponding to the
first half-period can be seen.

The first step for a  $\B$ meson oscillation analysis is the selection of
final states suitable for the study. The choice of the selection criterion
determines also the strategy for the tagging
of the meson flavour at decay time. Then, the flavour at production time
is tagged, to give the global mistag probability.
Finally, the proper time is reconstructed for each meson candidate, 
and the oscillation is studied by means of a likelihood fit to the 
distributions of decays tagged as mixed or unmixed.
\begin{figure}[t]
  \vspace{4.5cm}  \includegraphics{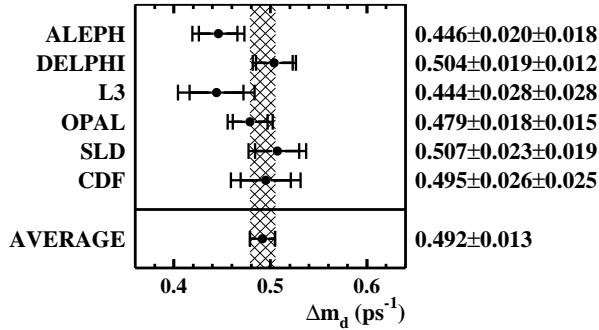}
  \caption{\it
    Measurements of $\Dmd$ from the LEP experiments, SLD and CDF,
    with the average from high-energy colliders.
    \label{fig.dmd} }
\end{figure}

Several measurements of the $\Bzero$ oscillations frequency have been
produced by the LEP experiments, SLD and CDF. A variety of 
selection methods have been used, offering different advantages
in terms of statistics, signal purity and control of the systematic uncertainties.
The average per experiment, and the global average are shown in
Fig.~\ref{fig.dmd}.
All the analyses rely on the simulation to some extent, and therefore are
affected by uncertainties in the physics processes that are simulated.
The results are adjusted to a common set of input parameters (\eg~b hadron
lifetimes and production fractions) and then averaged, deriving the
result of Fig.~\ref{fig.dmd} and the following values for the b hadron
production fractions:
\begin{equation}
f_{\Bzero , \Bplus} = (40.0 \pm 1.0) \% \ \ \  \ \ 
f_{\Bs} = (9.7 \pm 1.2) \%              \ \ \ \ \ 
f_{\mathrm{baryon}} = (10.3 \pm 1.7) \%  \ .
\label{eq:fracresult}
\end{equation}

In the case of $\Bs$ oscillations, the analyses currently available have not
been able to resolve the oscillation and produce a measurement of the 
frequency; only certain ranges of frequencies have been excluded. Combining such
excluded ranges is not straightforward, and a specific method, the 
{\em amplitude method}, was introduced for this purpose\cite{hgm}.
In the likelihood fit to 
the proper time distribution of decays tagged as mixed or unmixed, the 
frequency of the oscillation is not taken to be the free parameter,
but it is instead fixed to a ``test'' value $\omega$. An auxiliary 
parameter, the  amplitude $\amp$ 
of the oscillating term is introduced, and left free in the fit.
The proper time distributions for mixed and unmixed decay, prior to 
convolution with the experimental resolution, are therefore written as
\begin{equation}
  {\cal P}(t) = 
  \frac{\Gamma e^{-\Gamma t}}{2} \, [1 \pm \amp \cos (\omega\, t)] \ ,
\end{equation}
with $\omega$ the test frequency and $\amp$ the only free parameter. 
When the test frequency is much smaller than the true frequency 
(\mbox{$\omega \ll \Dms$}) the expected value for the amplitude is 
\mbox{$\amp = 0$}.
At the true frequency (\mbox{$\omega = \Dms$}) the expectation is 
\mbox{$\amp = 1$}.
All the values of the test frequency $\omega$ for which 
\mbox{$\amp + 1.645 \sigma_\amp < 1$} are excluded at $95\%$~C.L.
When $\omega$ approaches or exceeds the true frequency $\Dms$,
the shape of $\amp(\omega)$ depends on the details of the analysis
and can be calculated analytically in simple cases\cite{noi}.
The amplitude has well-behaved errors, and 
different measurements can be combined in a straightforward way, 
by averaging the amplitude measured at different test frequencies.
The excluded range is derived from the combined amplitude scan.

\begin{figure}[t]
  \vspace{7cm}  \includegraphics{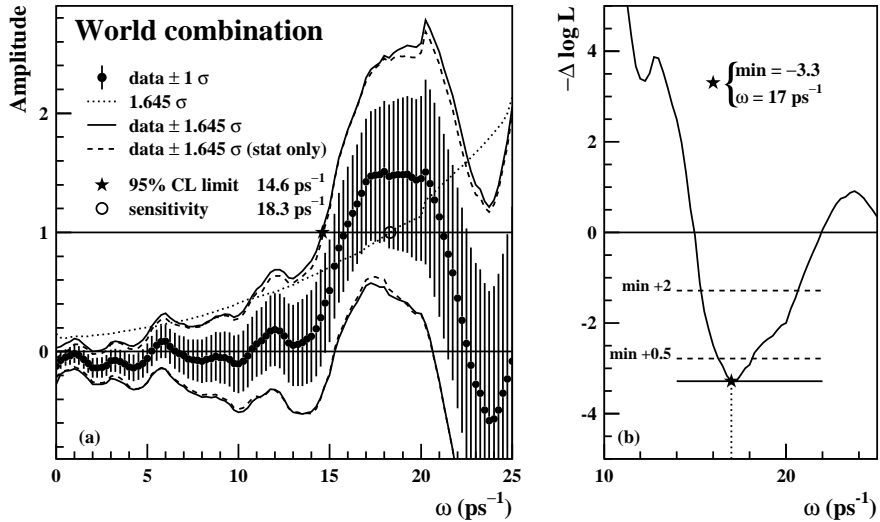}
  \caption{\it
    (a) Combined amplitude spectrum as a function of test frequency $\omega$.
   (b) Log-likelihood profile as derived from 
the amplitude spectrum (world combination). The dashed lines would represent the $1-2\ \sigma$ levels, 
if the likelihood was parabolic in a range wide
enough around the minimum.
     \label{fig:dms} }
\end{figure}

At present the world combination is dominated at high frequency 
by the analyses of ALEPH and SLD. The amplitude spectrum, 
with statistical and systematic errors is shown in Fig.~\ref{fig:dms}a.
A lower limit of \mbox{$\Dms>14.6 \ \ips$} is derived, while the
expected limit (sensitivity) is \mbox{$\Dms>18.3 \ \ips$}.
The difference is due to the positive amplitude values measured around 
\mbox{$17 \ \ips$}, compatible with one, as expected in the presence
of signal. The error on the amplitude at high frequency 
(\mbox{$\Dms\approx 20 \ \ips$}) is reduced by about a factor of two
compared to the world combination of Summer 1999, mostly because
of improvements in the analysis techniques. Some improvements
are still expected both from the the LEP experiments and SLD.

The amplitude spectrum can be translated to a log-likelihood profile,
referred to the asymptotic value for 
\mbox{$\Dms \to \infty$} (Fig.\ref{fig:dms}b). A minimum
is observed at $\Dms \approx 17\ \ips$. The deviation of the measured
amplitude from $\amp = 0$ around the likelihood minimum 
is about $2.5 \sigma$. Such a value cannot be used
to assess the probability of a fluctuation,
since it is chosen {\em a posteriori} among all the points
of the frequency scan performed. On the other hand because the amplitude
measurements at different frequencies are correlated, the probability
of observing a minimum as or more incompatible
with the hypothesis of background than the one found in the data,
needs to be estimated with toy experiments\cite{noi}. 
Such a probability is found to be about $3\%$.

\begin{figure}[tb!]
 \vspace{6.4cm}
  \includegraphics{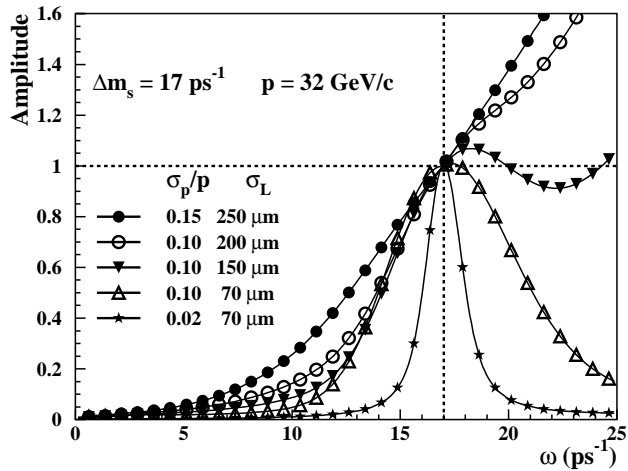}
 \caption{\it
   Expected amplitude shape for a true frequency $\Dms = 17 \ \ips$, monochromatic $\Bs$ mesons of $p=32\ \gevc$ and different values of momentum and decay length resolutions (taken to be Gaussian).
    \label{fig:ampl} }
\end{figure}

An interesting issue is to which extent the observation is compatible
with the hypothesis of signal. This cannot be assessed quantitatively in
 a non-trivial way. In Fig.~\ref{fig:ampl} the expected amplitude shapes,
calculated analytically\cite{noi},
are shown for the simple case of monochromatic  $\Bs$ mesons of  $p=32\ \gevc$
which oscillates with a frequency of  $17 \ \ips$, with different
(Gaussian) resolutions in momentum and decay length. The shapes 
are largely different; the only solid features are that the expectation is
$\amp=1$ at the true frequency and $\amp=0$ far below the true frequency. 
In the world combination many analyses contribute, which have widely different
momentum and decay length resolution. In the most
sensitive analyses, even, each event enters with its specific
estimated resolutions, therefore contributing 
with a different expected amplitude shape.
Calculating the expected amplitude shape for the world combination 
in the hypothesis of signal is therefore, at the moment, completely impractical.
It can certainly be stated, however, that qualitatively the shape observed
in Fig.~\ref{fig:dms}a is compatible with the hypothesis of a signal
at $\Dms\approx 17\ \ips$.

Indirect constraints on $\Dms$ can be derived, within the 
Standard Model framework, from other physics quantities.
Measurements of charmless b decays, CP
violation in the kaon system, and $\B$ meson oscillations
can all be translated, with nontrivial theoretical input,
to constraints on the ($\rho,\eta$) parameters\cite{stocchi}, and combined.
If the limit on $\Dms$
is removed from the fit, a probability
density function can be extracted from the other measurements.
The preferred value is
\mbox{$\Dms = 14.9 ^{+4.0} _{-3.6} \ \ips$}, perfectly compatible
with the indication observed in the combination of $\Dms$ analyses.
The present world average of the width difference in the $\Bs$ 
system presented in Section~\ref{sec.life},
can be also translated to a value
for the oscillation frequency, using the prediction of NLO+lattice
calculations\cite{beneke_dgs} for the ratio $\Delta \Gamma_s / \Dms$. 
That gives a mild constraint on the $\Bs$ oscillation frequency,
\mbox{$\Dms = 29 \, ^{+16} _{-21}\,  \ips$}, compatible
with the previous result.

\section{Conclusions}

Z factories have given a major contribution
to the knowledge of b hadron physics over the past decade.
Today asymmetric B factories are pushing forward our 
knowledge of $\Bzero/\Bplus$ physics, and of CP violation
in the b sector.

In this report, results on b hadron lifetimes have been
reviewed, together with measurements of $\aVcb$ and $\aVub$
from studies of semileptonic b decays,
measurements of the $\Bzero$ oscillation frequency and
limits on the $\Bs$ oscillation frequency. 

A deviation of about  $2.5\sigma$ from $\amp = 0$ is found around $17\ \ips$ 
in the $\Bs$ oscillation frequency scan, qualitatively compatible
with an oscillation signal. Improvements are still expected
in some LEP and SLD analyses in the coming months, which might help clarifying
whether or not the effect observed is evidence for a signal.

\section{Acknowledgements}
It is a pleasure to thank the organizers of the conference for the 
interesting meeting. The averages presented have been provided
by the Working Groups on Electroweak Heavy Flavour Physics, b lifetimes,
$\Vcb$, $\Vub$ and B Oscillations.

\end{document}